\documentclass[12pt]{iopart}

\usepackage{float}
\usepackage{epsfig}
\usepackage[latin1]{inputenc}
\usepackage{theorem}
\usepackage{psfrag}
\usepackage{amsfonts}
\usepackage{amssymb}

\theoremstyle{plain}

\newtheorem{proposition}{Proposition}

\newtheorem{definition}{Definition}

\newenvironment{proof}[1][Proof]{\textbf{Proof} }{\
\rule{0.5em}{0.5em}}

\newcommand{\qu}[1]{\ensuremath{|#1\rangle}}

\newcommand{\ip}[2]{\ensuremath{\langle #1|#2\rangle}}
\newcommand{\op}[1]{\ensuremath{|#1\rangle\langle #1|}}
\newcommand{\tra}[2][]{\ensuremath{\textrm{tr}_{#1} \left[#2 \right] }}
\newcommand{\mce}{\ensuremath{{\cal{E}}}}
\newcommand{\mch}{\ensuremath{{\cal{H}}}}
\newcommand{\mcc}{\ensuremath{{\cal{C}}}}

\newcommand{\rl}{\ensuremath{\overline{\rho}}}

\newcommand{\mcs}{\ensuremath{{\cal{S}}}}

\newcommand{\mcp}{\ensuremath{{\cal{P}}}}

\def\one{{\mathchoice{\rm 1\mskip-4mu l}{\rm 1\mskip-4mu l}{\rm 1\mskip-4.5mu l}{\rm
1\mskip-5mu l}}}

 \newcommand{\E}{{\mathcal{E}}}
 
 \newcommand{\G}{{\mathcal{G}}}
\renewcommand{\H}{{\mathcal{H}}}

\renewcommand{\P}{{\mathcal{P}}}

 \newcommand{\V}{{\mathcal{V}}}
 
\begin{document}

\title{Quantum states characterization for the zero-error capacity}

\author{Rex A C Medeiros$^{\dag,\ddag,1,2}$ Romain Alléaume$^{\dag,2}$, Gérard Cohen$^{\dag,3}$ and 
Francisco M. de Assis$^{\ddag,4}$ }

\address{
~\\
$^\dag$ Département Informatique et Réseaux,\\
 École Nationale Supérieure des Télécommunications \\
 46 rue Barrault, F-75634, Paris Cedex 13, France\\
 ~
}

\address{
$^\ddag$ Departamento de En\-ge\-nha\-ria Elétrica \\
 Universidade Federal de Campina Grande \\
 Av. Apr\'{i}gio Veloso, 882, Bodocong\'{o}\\
Campina Grande-PB, 58109-970, Brazil \\
~
}

\ead{$^1$ rex.medeiros@enst.fr}
\ead{$^2$ romain.alleaume@enst.fr}
\ead{$^3$ gerard.cohen@enst.fr}
\ead{$^4$ fmarcos@dee.ufcg.edu.br}

\begin{abstract}
The zero-error capacity of quantum channels was defined as the least upper bound of rates at which classical information can be transmitted through a quantum channel with probability of error equal to zero. This paper investigates some properties of input states and measurements used to attain the quantum zero-error capacity. We start by reformulating the problem of finding the zero-error capacity in the language of graph theory. This alternative definition is used to prove that the zero-error capacity of any quantum channel can be reached by using tensor products of pure states as channel inputs, and projective measurements in the channel output. We conclude by presenting an example that illustrates our results. 
\end{abstract}

\pacs{03.67.-a, 03.67.Hk}
\submitto{\JPA}
\maketitle

\section{Introduction}

Classical and quantum information theory~\cite{Cover:91,BS:98} usually look for asymptotic solutions to information treatment and transmission problems. For example, the Shannon's coding theorem guarantees the existence of a channel capacity $C$ such that for any rate $R$ approaching $C$ there exist a sequence of codes for which the probability of error goes asymptotically to zero. A zero-error probability approach for information transmission through noisy channel was introduced by Shannon in 1956~\cite{Sha:56}. Given a discrete memoryless channel, it was defined a capacity for transmitting information with an error probability equal to zero. The so called zero-error information theory~\cite{KoOr:98} found applications in areas like graph theory, combinatorics, and computer science.

More recently, the zero-error capacity  of  quantum channels was defined as the least upper bound of rates at which classical information can be transmitted through a quantum channel with error probability equal to zero~\cite{MeAs:05a}. Some  results followed the definition. For example, it was shown that the zero-error capacity of any quantum channel is upper bounded by the HSW capacity~\cite{MeAs:04d}. 

Because of the direct relation with graph theory, the quantum zero-error capacity should have connections with several areas of quantum information and computation, like quantum error-correction codes~\cite{ZR97c}, quantum noiseless subsystems~\cite{Zan01,CK:05}, faut-tolerant quantum computation~\cite{KBLW01a}, graph states~\cite{HEB04}, and quantum computation complexity.

In this paper we give an alternative definition for the zero-error capacity of quantum channels in terms of graph theory. Also, we present new results concerning quantum states attaining the quantum channel capacity. Particulary, we show that non-adjacent states live into orthogonal Hilbert subspaces, and that non-adjacent states are orthogonal. Our main result asserts that the quantum zero-error capacity can be reached by using only pure states. In addition, we prove that general POVM measurements are not required: for a given quantum channel, it is always possible to find a von Neumann measurement attaining the quantum zero-error capacity. A mathematically motivated example is given to illustrate our results.

The rest of this paper is structured as follows. Section~\ref{sec:back} recalls some definitions concerning the  zero-error capacity of a quantum channel. Section 3 reformulates the problem of finding the quantum zero-error capacity into the graph language. This alternative definition is used in Sec.~\ref{sec:charact} to study the behavior of input states. Section~\ref{sec:measurements} discusses about measurements reaching the quantum zero-error capacity, and Sec.~\ref{sec:ex} illustrates the results with an example.  Finally, Sec.~\ref{sec:conclu} presents the conclusions and discusses further works.

\section{Background\label{sec:back}}

We review some important definitions. Consider a $d-$dimensional quantum channel $\E \equiv \{E_a\}$ and a subset $\mcs$ of input states, and let $\rho_i \in \mcs$. We denote $\sigma_i = \mce(\rho_i)$ the received quantum state when $\rho_i$ is transmitted through the quantum channel. Define a POVM  $\{M_j\}$, where $\sum_j M_j =\one$. For convenience, we call Alice the sender and Bob the recipient. If $p(j|i)$ denotes the probability of Bob gets the outcome $j$ given that Alice sent the state $\rho_i$, then, $ p(j|i) = \tra{\sigma_{i} M_{j}}$.

By analogy with classical information theory~\cite{Sha:56}, the zero-error capacity of a quantum channel is defined for product states. A product of any $n$ input states will be called an input quantum codeword, $\rl_i = \rho_{i_1} \otimes \dots \otimes \rho_{i_n}$, belonging to a $d^n$-dimensional Hilbert space $\mch^n$. A mapping of $K$ classical messages (which we may take to be the integers $1,\dots,K$) into a subset of input quantum codewords will be called a quantum block code of length $n$. Thus, $\frac{1}{n}\log K$ will be the rate for this code. A piece of $n$ output indices obtained from measurements performed by means of a POVM $\{M_1,\dots,M_m\}$ will be called an output word, $w \in \{1,\dots,m\}^n$. 

A decoding scheme for a quantum block code of length $n$ is a function that univocally associates each output word with integers 1 to $K$ representing classical messages. The probability of error for this code is greater than zero if the decoding system identifies a different message from the message sent.

\begin{figure}[h]
\centering
\psfrag{ie1k}[l]{\small $i\in\{1,\dots,K\}$}
\psfrag{spsi}[l]{\small $\mcs = \{\qu{\psi_i}\}$}
\psfrag{psin}[l]{\small $\qu{\psi_i}^{\otimes n}$}
\psfrag{Ep}[l]{\small $\E(\cdot)$}
\psfrag{POVM}[l]{\footnotesize POVM}
\psfrag{Mi}[l]{\small $\{M_i\}_{i=1}^m$}
\psfrag{we1m}[l]{\small $w\in\{1,\dots,m\}^n$}
\psfrag{je1k}[l]{\small $\tilde{\imath}\in\{1,\dots,K\}$}
\psfrag{comment}[l]{\small $p(\tilde{\imath}|i)$: discrete memoryless classical channel}
\includegraphics[scale=1]{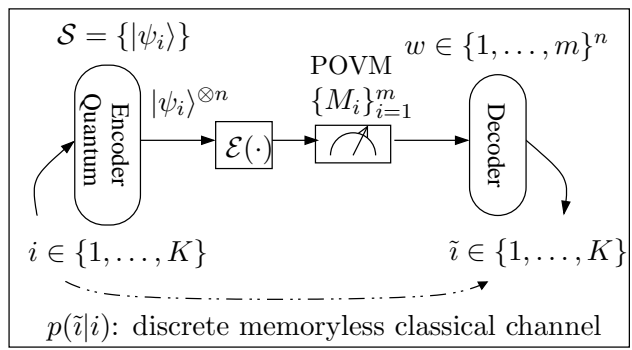} 
\caption{\label{fig:system} 
General representation of a quantum  zero-error communication system.}
\end{figure}

Figure~\ref{fig:system} illustrates a system that makes use of a quantum channel to transmit classical information with a probability of error equal to zero. Initially, Alice chooses a message $i$ from a set of $K$ classical messages. Based on the message $i$ and on the structure of a quantum block code of length $n$, the quantum  encoder prepares a $n$-tensor product of quantum states which is sent through the quantum channel $\mce(\cdot)$. In the reception, Bob performs a POVM measurement in order to obtain an output word $w$. The decision system, called decoder, should associate the output word $w$ with a message $\tilde{\imath}$. In a zero-error context it is required $\tilde{\imath} = i$ with probability one.

\normalsize

\begin{definition}
\label{def:qzec}
Let $\mce(\cdot)$ be a trace-preserving quantum map representing a noisy quantum channel. The zero-error capacity of~$\mce(\cdot)$, denoted by $C^{(0)}(\mce)$, is the least upper bound of achievable rates with probability of error equal to zero. That is,
\begin{equation}
\label{eq:qzec}
  C^{(0)}({\mce}) = \sup_{n} \frac{1}{n}\log K(n),
\end{equation}
where $K(n)$ stands for the maximum number of classical messages that the system can transmit without error, when a quantum block code of length $n$ is used.
\end{definition}

A canonical method for calculating the supremum in the Eq.~(\ref{eq:qzec}) involves a search on all possible input state subsets $\mcs$ and  POVMs $\P$. Given a particular $(\mcs,\P)$, $\mcs = \{\rho_{1},\dots,\rho_{l}\}$, $\P = \{M_{1},\dots,M_{m}\}$, and supposing a memoryless quantum channel, one may define a classical discrete memoryless channel (DMC) as follows. Take indexes $j$ of $\rho_{j}$ and $k$ of $M_{k}$ as input and output alphabets, respectively. The transition matrix will be a $||\mcs||\times||\P||$ matrix given by $T=[p(k|j)]$, where 
\begin{equation}
\label{eq:transitionmatrix}
p(k|j)=\tra{\E(\rho_{j})M_{j}}.
\end{equation}
Clearly, this classical equivalent channel has a zero-error capacity. Then, the zero-error error capacity of the quantum channel will be the maximum of these capacities over all possibles $(\mcs,\P)$. 

\begin{definition} 
\label{def:optimum}
An optimum $(\mcs,\mcp)$ for a quantum channel $\mcc$ is composed of a set $\mcs = \{\rho_i\}$ and a POVM $\mcp = \{M_j\}$ for which the zero-error capacity is reached.
\end{definition}

Next we recall the  definition of non-adjacent states.

\begin{definition}
Two quantum states $\rho_1$ and $\rho_2$ are said to be non-adjacent with relation to a POVM $\mcp = \{M_j\}_{j=1}^{m}$ if $A_1 \cap A_2 =\oslash$, where
$$
A_k = \{ j \in \{1,\dots,m\}; \; \tra{\mce(\rho_k)M_{j}} > 0\}; \; k = 1,2.
$$
\end{definition}

We proved a necessary and sufficient condition for which a quantum channel has zero-error capacity greater than zero:

\begin{proposition}[\cite{MeAs:05a}]
\label{pro:cond}
The zero-error capacity of a quantum channel is greater than zero if and only if there exist a subset $\mcs=\{\rho_i\}_{i=1}^{l}$ and a POVM $\mcp = \{M_j\}_{j=1}^{m}$ for which at least two states in $\mcs$ are non-adjacents with relation to the POVM $\P$.
\end{proposition}

\section{Relation with graph theory}

Given a classical discrete memoryless channel, two input symbols  are adjacent if there is an output symbol which can be caused by either of these two. From such channels, we may construct a graph $G$ by taking as many vertices as the number of input symbols, and connecting two vertices if the corresponding input symbols are non-adjacent. Shannon~\cite{Sha:56} showed that the zero-error capacity of the DMC is given by
$$
 C = \sup_n  \frac{1}{n} \log \omega\left(G^n\right) ,
$$
where $\omega (G)$ is the clique number of the graph $G$ and $G^n$ is the $n-$product graph of $G$.

The problem of finding the zero-error capacity of a quantum channel is straightforwardly reformulated in the language of graph theory. Given a subset of input states $\mcs_{(i)}$ and a POVM $\P_{(i)}$, we can construct a characteristic  graph $\G_{(i)}$ as follows. Take as many vertices as $||\mcs_{(i)}||$ and connect two vertices if the corresponding input states in $\mcs_{(i)}$ are non-adjacents for the POVM $\P_{(i)}$. 

\begin{definition}[Alternative definition]
\label{def:equiv}
The zero-error capacity of the quantum channel is given by
\begin{equation}
\label{eq:clique}
C^{(0)}({\mce}) =  \sup_{(\mcs_{(i)},\P_{(i)})} \sup_n  \frac{1}{n} \log \omega\left(\G_{(i)}^n\right),
\end{equation}
where $\omega (\G)$ is the clique number of the graph $\G$ and $\G_{(i)}^n$ is the $n-$product graph of $\G_{(i)}$. 
\end{definition}

It is easy to see that the supremum in Eq.~(\ref{eq:clique}) is achieved for the optimum $(\mcs,\P)$. Moreover, the characteristic graph we construct from the transition matrix defined by Eq.~(\ref{eq:transitionmatrix}) is identical to $\G_{(i)}$. We use this alternative definition to prove further results.

\section{Characterizing input states\label{sec:charact}}

It is known that finding the clique number of a graph (and consequently que zero-error capacity) is a NP-complete problem~\cite{Bollo:98}. One might expect that calculating the zero error-capacity of quantum channels is a more difficult task. For such channels, this process involves a search for the optimum $(\mcs,\P)$. For example, a priori the subset $\mcs$ may contain any kind of quantum states. The results presented in this section aim to reduce the search space of operators in $\mcs$. Particularly, we show that it is only needed to consider pure states to attain the supremum in~Eq.(\ref{eq:clique}). 

Proposition below relates orthogonality of output states and adjacency.

\begin{proposition}\label{teo:orto}
For a quantum channel $\E \equiv \{E_a\}$, two input states $\rho_1,\rho_2 \in \mcs$ are non-adjacent for a given POVM $\mcp = \{M_1,\dots,M_m\}$ if and only if $\E(\cdot)$ takes $\rho_1$  and $\rho_2$ into orthogonal subspaces.
\end{proposition}

More specifically, Proposition~\ref{teo:orto} asserts that if $\rho_1$ and $\rho_2$ are non-adjacent, then their images $\E(\rho_1)$ and $\E(\rho_2)$ are entirely  inside orthogonal Hilbert subspaces.  At  first glance this seems to be an obvious result. However, remember that $\E(\rho_i)$ may be  mixed states and it is important to know in which subspace each of them lives.

\begin{proof}
Given a complete set of POVM operators $\mcp = \{M_1,\dots,M_m\}$, a POVM measurement apparatus can be viewed as a black box that outputs a number from $1$ to $m$ when an unknown quantum state is measured.

Suppose that $\rho_1$ and $\rho_2$ are non-adjacent quantum input states. For integers $k,l$ satisfying $k+l\leq m$, we can  always reorder the POVM indexes so that $\mcp = \{M_1,\dots,M_k,\dots,M_{k+l},\dots,M_m\}$and
$$
\textrm{Prob } [ i \; | \;  \rho_1 \textrm{ was sent }] 
\begin{cases}
> 0 \; \forall \; i=1,\dots,k\\
= 0 \; \textrm { otherwise}
\end{cases}
$$
and
$$
\textrm{Prob } [ i \; | \;  \rho_2 \textrm{ was sent }] 
\begin{cases}
> 0 \; \forall \; i=k+1,\dots,k+l \\
=   0 \;  \textrm{ otherwise}.
\end{cases}
$$

This scenario is  explained in Fig.~\ref{fig:povm}. On the left side we put the states $\rho_i$, and all POVM elements on the right side. Next we draw a line from $\rho_i$ to $M_j$ if $\textrm{Prob } [ \textrm{get output } j \; | \;  \rho_i \textrm{ was sent }]  = \tra{\E(\rho_i)M_j}> 0$.


\begin{figure}[hbt]
\centering
\psfrag{p1}[c]{$\rho_1$}
\psfrag{p2}[c]{$\rho_2$}
\psfrag{1}[cl]{$M_1$}
\psfrag{2}[cl]{$M_2$}
\psfrag{km1}[cl]{$M_{k-1}$}
\psfrag{k}[cl]{$M_{k}$}
\psfrag{kma1}[cl]{$M_{k+1}$}
\psfrag{kma2}[cl]{$M_{k+2}$}
\psfrag{kml}[cl]{$M_{k+l}$}
\psfrag{kmlm1}[cl]{$M_{k+l-1}$}
\psfrag{km}[cl]{$M_{m}$}
\psfrag{r}[c]{$\vdots$}
\includegraphics[scale=.9]{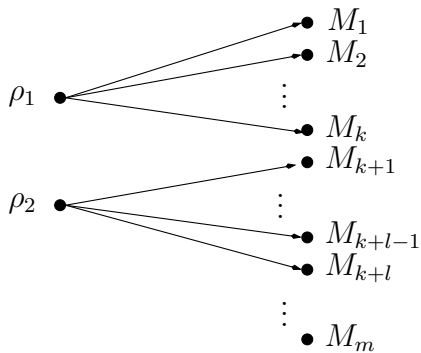} 
\caption{\label{fig:povm} Two non-adjacent quantum states for the POVM $\mcp$. The same method is employed to construct the classical equivalent discrete memoryless channel (DMC) used to calculate the zero-error capacity of quantum channels (see~\cite{MeAs:05a}).}
\end{figure}

It is possible to build a new POVM containing only two elements $\{M^{(1)},M^{(2)}\}$ as

\begin{equation}
M^{(1)} = \sum_{i=1}^{k} M_i  \qquad \textrm{and} \qquad M^{(2)} = \sum_{i=k+1}^{m} M_i
\end{equation}
for which
\begin{eqnarray*}
\textrm{Prob } [ \textrm{get output } (1) \; | \;  \rho_1 \textrm{ was sent }] &=& 1 \\
\textrm{Prob } [ \textrm{get output } (2) \; | \;  \rho_2 \textrm{ was sent }] &=& 1,
\end{eqnarray*}
or equivalently,
\begin{eqnarray*}
\tra{\E(\rho_1)M^{(1)}} &=& 1 \\
\tra{\E(\rho_2)M^{(2)}} &=& 1.
\end{eqnarray*}

For the ``if'' part it is sufficient to demonstrate that $M^{(1)}$ and $M^{(2)}$ are orthogonal projectors. Note that $M^{(1)} + M^{(2)} = \one$. Hence, if $M^{(1)}$ is a projector, then $M^{(2)}$ is its orthogonal complement.

Let $\E(\rho_1) = \sum_a E_a \rho_1 E_a^\dagger$ be the output state when $\rho_1$ is sent through the quantum channel. The spectral decomposition of $\E(\rho_1)$ gives us
$$
    \E(\rho_1) = \sum_i \alpha_i^{(1)} \op{a_i},
$$
for an orthonormal base $\qu{a_i}$ and positive numbers $\alpha_i^{(1)}$, $\sum_i \alpha_i^{(1)} = 1$. Then, verifying $\tra{\E(\rho_1)M_1^{(1)}} = 1 $ implies
\begin{eqnarray*}
\tra{M^{(1)}\sum_i \alpha_i^{(1)} \op{a_i}} &=& \sum_i \alpha_i^{(1)} \ip{a_i}{M^{(1)}|a_i} \\
&=& \sum_i \alpha_i^{(1)} \\
&=& 1.
\end{eqnarray*}
Notice that $M^{(1)}$ is a positive matrix satisfying $M^{(1)} \leq \one$. From this we conclude that $\ip{a_i}{M^{(1)}|a_i} = 1 \; \forall \; i $ such that $\qu{a_i}$ is in the support of $  {\E(\rho_1)}$. Finally, we can write $M^{(1)}$ as
$$
    M^{(1)} = \sum_{\{i:\qu{a_i} \in \sup {\E(\rho_1)}  \} }\op{a_i},
$$
which is a projector on the subspace spanned by the eigenvectors of $\E(\rho_1)$ with nonzero eigenvalues.

Conversely, let $\E$ be a quantum channel that take $\rho_1$ and $\rho_2$ into orthogonal subspaces. If  $M^{(1)}$ and $M^{(2)}$ are projectors over these subspaces, then
$$
    \tra{\E(\rho_1)M^{(1)}} = 1 \quad \Rightarrow \quad \tra{\E(\rho_1)M^{(2)}} = 0
$$
and
$$
    \tra{\E(\rho_2)M^{(2)}} = 1 \quad \Rightarrow \quad \tra{\E(\rho_2)M^{(1)}} = 0,
$$
and the result follows.
\end{proof}

We recall the definition of the Holevo-Schumacher-Westmoreland's classical capacity for a quantum channel~\cite{Hol:98,SW:97}: 
{\small
$$
 C_{1,\infty}(\mce) \equiv \max_{\{p_i,\rho_i\}} \left[ S\left(\mce\left(\sum_i p_i\rho_i\right)\right) - \sum_i p_i  S(\mce(\rho_i))\right].
$$
}
A very interesting result about this capacity claims that the maximum is reached by using only pure states, i.e., we need only consider states like $\rho_i =  \op{v_i}$  in the input of the channel.

For the quantum zero-error capacity (QZEC), we have an analogous result:

\begin{proposition}\label{teo:puros}
The QZEC of quantum channels is calculated by using an optimum map  $(\mcs,\mcp)$, where the set $\mcs$ is composed only by pure quantum states, i.e., $\mcs = \{\rho_i = \op{v_i}\}$.
\end{proposition}

\begin{proof} 
Consider a quantum channel represented by a trace-preserving linear map, $\mce(\cdot)$, with operation elements $\{E_a\}$. Suppose $(\mcs,\mcp)$ is an optimum map, with $\mcs = \{\rho_1,\dots,\rho_l\}$ and $\mcp = \{M_1,\dots,M_m\}$, and each state $\rho_i$ may be a mixed state. We call $\G $ the  characteristic graph associated with $(\mcs,\mcp)$. To demonstrate the proposition, we show that it is always possible to obtain a subset $\mcs'$ from $\mcs$, such that  $\mcs'$ contains only pure states and $(\mcs',\mcp'=\mcp)$ is also optimum.

Let $\rho_i \in \mcs$, $\rho_i = \sum_v \lambda_{v_i} \op{v_i}$ be an input quantum state. Then, the output of the channel when $\rho_i$ is transmitted is given by

\begin{eqnarray}
  \mce(\rho_i) &=& \sum_a E_a \rho_i E_a^\dagger \nonumber\\
   &=& \sum_a E_a \left[ \sum_v \lambda_{v_i} \op{v_i} \right] E_a^\dagger \nonumber\\
   &=& \sum_a \sum_v E_a \lambda_{v_i} \op{v_i} E_a^\dagger.
\end{eqnarray}

By using the POVM $\mcp$, the probability of measuring $j$ given that the quantum state $\rho_i$ was sent is 

\begin{eqnarray}
  p(j|i) &=& \tra{\mce(\rho_i) M_j} \nonumber \\
  &=& \tra{\left(\sum_a \sum_v E_a \lambda_{v_i} \op{v_i} E_a^\dagger \right) M_j} \nonumber \\
  &=& \sum_v \lambda_{v_i} \tra{\left( \sum_a E_a \op{v_i} E_a \right) M_j}.
\label{eq:traco}
\end{eqnarray}

Note that in the equation above, $\tra{\cdot}$ is always greater than or equal to zero and $0< \lambda_{v_i}\leq 1$. It represents the probability of getting output $j$ given that the pure state $\qu{v_i}$ was sent through the quantum channel. If we replace the mixed states $\rho_i$ by any pure state $\qu{v_i}$ in the support of $\rho_i$, the cardinality of the subset $A_i$ (see Def. 3) never increases. To see this, let $M_k$ be an POVM element so that $\tra{\E(\rho_i)M_k} = 0$. From Eq.~(\ref{eq:traco}),
\begin{eqnarray}
\tra{\E(\rho_i)M_k} &=& \sum_v \lambda_{v_i}  \tra{\left( \sum_a E_a \op{v_i} E_a \right) M_k} \nonumber\\
&=& 0
\end{eqnarray}
implies $ \tra{\left( \sum_a E_a \op{v_i} E_a \right) M_k} = 0$ for all pure states $\qu{v_i}$ in the support of $\rho_i$. Now define a new set $\mcs'$ by replacing each mixed state $\rho_i \in \mcs$ with a pure state $\qu{v_i} \in \sup \rho_i$. The number of non-adjacent states in $\mcs'$ is at least that of $\mcs$.  A larger number of non-adjacency leads to a more connected characteristic  graph. For any graph $G$, and in particular for the characteristic graph, it is well known that adding edges never decreases (and may increase) the clique number~\cite{Bollo:98}, and according to Eq.~(\ref{eq:clique}) this may not reduce the zero-error capacity of the quantum channel.

Finally, we may always find a set $\mcs' = \{\rho'_1,\dots,\rho'_l\}$, where $\rho'_i = \op{v_i} \in \sup \rho_i$  and $(\mcs',\mcp)$ is also optimum.

\end{proof}

The proposition~\ref{teo:puros} allow us to prove the next result considering only pure states:

\begin{proposition}
Let $\qu{v_1}$ e $\qu{v_2}$ be two non-adjacent states. Then, $\ip{v_1}{v_2}=0$.
\end{proposition}

~
\begin{proof}
To prove the proposition, we make use of a distance measure for quantum states called trace distance. The trace distance between $\sigma_1$ and $\sigma_2$ is given by
$$
D(\sigma_1,\sigma_2) = \frac{1}{2} \textrm{tr }\left|\sigma_1 - \sigma_2 \right|.
$$
Note that the trace distance is maximum and equal to one if, and only if, $
\sigma_1$ and $\sigma_2$ have orthogonal supports. 

Proposition~\ref{teo:orto} guarantees  that if $\qu{v_1}$ and $\qu{v_2}$ are non-adjacent, then $\E(\qu{v_1})$ and $\E(\qu{v_2})$ have orthogonal supports. Because we assumed  $\qu{v_1}$ and $\qu{v_2}$ non-adjacent, we have
$$
D(\E(\qu{v_1}),\E(\qu{v_2})) = 1.
$$
It is easy to show that quantum channels $\E \equiv \{E_a\}$ are contractive~\cite[pp. 406]{NC:2000}, i.e., $D(\qu{v_1},\qu{v_2}) \ge D(\E(\qu{v_1}),\E(\qu{v_2}))$. The result now follows:
\begin{equation}
1 \ge D(\qu{v_1},\qu{v_2}) \ge D(\E(\qu{v_1}),\E(\qu{v_2})) =1,  \\
\end{equation}
which means that  $D(\qu{v_1},\qu{v_2}) = 1$ and $\qu{v_1}$ are orthogonal to  $\qu{v_2}$ .
\end{proof}

Consider a qubit channel and an orthonormal basis for the 2-dimensional Hilbert space. Our results allow for the analysis  of  such channels in a zero-error context: either the zero-error capacity is equal to one bit per use or to zero. This is because these channels have at most two non-adjacent input states. If we take any subset $\mcs$ containing $n$ states, $n-2$ states will be adjacent with at least one of the others two. 

For a quantum channel in a $d-$dimensional Hilbert space, the canonical method presented in Sec.~\ref{sec:back} can be improved. The search for the subset $\mcs$ should start by taking sets of orthogonal pure states. Evidently, adjacent states can be added to the initial set if they contribute to increase the clique number in Eq.~(\ref{eq:clique}).

\section{POVM measurements attaining the zero-error capacity\label{sec:measurements}}

As pointed in the Sec.~\ref{sec:charact}, the problem of calculating the quantum zero-error capacity should be, in general, more difficult than finding the classical zero-error capacity. This fact can be understood  by analyzing  Eq.~(\ref{eq:clique}). Each choice of $(\mcs_{(i)},\P_{(i)})$ gives rise to a classical channel. Therefore, the quantum zero-error capacity can be interpreted as the supremum of the classical zero-error capacity of such equivalent channels over all possible $(\mcs_{(i)},\P_{(i)})$. Hence, a major issue is to restrict the global search space of operators in $\mcs$ and measurements $\P$. The main result in Sec.~\ref{sec:charact} claims that we only need to consider pure states in $\mcs$ to attain the zero-error capacity of a quantum channel.

Concerning the measurements, We have proven that we can restrict $\P$ to a projective measurement in order to attain the supremum in  Eq.~(\ref{eq:clique}).

\begin{proposition}\label{teo:povm}
The QZEC  can be calculated by using an optimum map  $(\mcs,\mcp)$, where $\P$ stands for a set of von Neumann operators $M_i$ with $\sum_i M_i = \one$.
\end{proposition}

\begin{proof}
First consider an optimum $(\mcs,\P)$, where $\mcs = \{\qu{v_i}\}$, $\P  = \{M_i\}$, and $M_i$ are positive operators satisfying $\sum_i M_i = \one$. Let $\H_i$ be the Hilbert space spanned by the states in the support of $\E(\qu{v_i})$. It is  known that $\qu{v_i}$ and $ \qu{v_j}$ are non-adjacent if and only if $\H_i \bot \H_j$.

Let $\G$ be the characteristic graph for the optimum $(\mcs,\P)$.  To demonstrate the result,  it is only need to show the existence of a Von Neumann measurement giving rise to the same characteristic graph.

Consider $\V_i$  an orthonormal basis of $\H_i$, and $\V$ an orthonormal  basis of $\H$, the whole Hilbert space of dimension $d$. From  $\G$, it \emph{ should} exist at least one orthonormal basis $\V$ of $\H$ , say $\V = \{\qu{\varphi_1},\dots,\qu{\varphi_d}\}$, with $\V_i \subset \V$, for which all (non-)adjacency relations in $\G$ are satisfied.

Now define a POVM $\P' = \{\op{\varphi_i}\}$. It is easy to see that $(\mcs,\mcp')$ gives rise to the same characteristic graph $\G$. 
\end{proof}

Note that each $\V_i$ is not necessarily composed by the vectors on the support of $\E(v_i)$. This is only true if all quantum states $\E(v_i)$ are mutually orthogonal.

\section{A non-trivial example: the pentagon\label{sec:ex}}

We discuss in this section an example of a quantum channel which has a non-trivial zero-error capacity.  By non-trivial we mean a channel whose characteristic graph for the optimum $(\mcs,\P)$ is neither a complete nor a empty graph, \emph{and} for which the supremum in Eq.~(\ref{eq:clique}) is attained  for $n>1$. Trivial examples of the quantum zero-error capacity can be found in~\cite{MeAs:05a}. The following example is mathematically motivated, and has not a physical meaning. However, it is interesting because the quantum channel we constructed gives rise to the pentagon as the characteristic graph for the optimum $(\mcs,\P)$. Historically, the zero-error capacity of the pentagon was studied by Shannon~\cite{Sha:56} in 1956, that gives  lower and upper bounds for it. Only in 1979, Lovász~\cite{Lov:79} gave an exact solution for this problem.

Let $\E(\cdot)$ be a quantum channel with Kraus operators $\{E_1,E_2,E_3\}$ given by 
$$
E_1 = 
\left( \begin{array}{ccccc}
  \alpha & 0 & 0 & 0 & \beta \\
  \alpha & \beta & 0 & 0 & 0 \\
  0 & \alpha & \beta & 0 & 0 \\
  0 & 0 & \alpha & \beta & 0 \\
  0 & 0 & 0 & \alpha & \beta
\end{array} \right),
\qquad 
E_2 = 
\left( \begin{array}{ccccc}
  \alpha & 0 & 0 & 0 & -\beta \\
  \alpha & -\beta & 0 & 0 & 0 \\
  0 & \alpha & -\beta & 0 & 0 \\
  0 & 0 & \alpha & -\beta & 0 \\
  0 & 0 & 0 & \alpha & -\beta
\end{array} \right),
$$
$$
E_3 = 
\left( \begin{array}{ccccc}
  \sqrt{1-4\alpha^2} & 0 & 0 & 0 & 0 \\
  0 & \sqrt{1-2\alpha^2-2\beta^2}& 0 & 0 & 0 \\
  0 & 0 & \sqrt{1-2\alpha^2-2\beta^2} & 0 & 0 \\
  0 & 0 & 0 & \sqrt{1-2\alpha^2-2\beta^2} & 0 \\
  0 & 0 & 0 & 0 & \sqrt{1-4\beta^2}
\end{array} \right),
$$
where $0\leq \alpha,\beta \leq 0,5$. It is easy to see that $\sum_a E_a^\dagger E_a = \one$, and $\E(\cdot)$ is a linear trace-preserving quantum map that represents a physical process.

\begin{figure}[h]
\centering
\psfrag{g1}[c]{$\qu{v_1}$}
\psfrag{g2}[c]{$\qu{v_3}$}
\psfrag{g3}[c]{$\qu{v_5}$}
\psfrag{g4}[c]{$\qu{v_2}$}
\psfrag{g5}[c]{$\qu{v_4}$}
\psfrag{1}[c]{$\qu{v_1}$}
\psfrag{2}[c]{$\qu{v_2}$}
\psfrag{3}[c]{$\qu{v_3}$}
\psfrag{4}[c]{$\qu{v_4}$}
\psfrag{5}[c]{$\qu{v_5}$}
\psfrag{p1}[l]{$M_1$}
\psfrag{p2}[l]{$M_2$}
\psfrag{p3}[l]{$M_3$}
\psfrag{p4}[l]{$M_4$}
\psfrag{p5}[l]{$M_5$}
\includegraphics[scale=1]{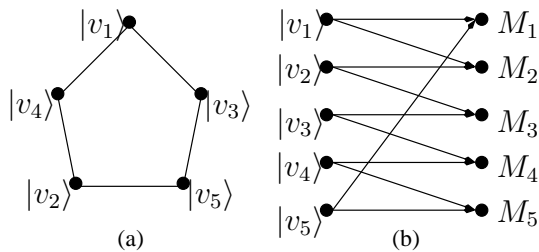} 
\caption{\label{fig:pentagono} 
Characteristic graph for the optimum $(\mcs,\P)$.
}
\end{figure}

The channel model was proposed in a way that the optimum $(\mcs,\P)$ is given by:
$$
 \mcs = \{\qu{v_1},\dots,\qu{v_5}\} \qquad \P = \{\op{v_i}\}_{i=1}^5,
$$
where $\qu{v_i}$ is the computation basis of the Hilbert space of dimension five. Once we have the optimum $(\mcs,\P)$, the  adjacency relationships are easily obtained by taking    
\begin{equation}
  \mathrm{Prob ~}(\mathrm{get~output  ~ } j| \; \qu{\psi_i} \mathrm{ ~was~ sent}) = \tr{\E(\op{v_i})M_j}; \quad i,j=1,\dots,5.
\end{equation}
The characteristic graph for the optimum $(\mcs,\P)$ is illustrated in Fig.~\ref{fig:pentagono}.

The Shannon capacity of this graph was calculated by Lovázs~\cite{Lov:79}, and it is attained for $n=2$:
$$
 C^{(0)}(\mathrm{pentagon}) =   \frac{1}{2} \log 5.
$$
Therefore, $ C^{(0)}(\mathrm{pentagon})$ is the maximum rate for which classical information can be transmitted through the quantum channel with a probability of error equal to zero. A quantum block code  of length two reaching the capacity is
\begin{eqnarray}
\rl_1 = \qu{v_0}\qu{v_0}, \quad \rl_2 = \qu{v_1}\qu{v_2}, \quad \rl_3 = \qu{v_2}\qu{v_4}\nonumber\\
\rl_4 = \qu{v_3}\qu{v_1}, \qquad \rl_5 = \qu{v_4}\qu{v_3}.
\end{eqnarray}

\section{Conclusions\label{sec:conclu}}
We have presented in this paper some results concerning the characterization of input states for the calculation of the zero-error capacity of quantum channels.

We initially showed that calculating the zero-error capacity of such channels is equivalent to finding the clique number of graph products. This first result is used to prove the main result of this paper: we have shown that the quantum zero-error capacity can be reached by using only pure input states.  In the literature, it was demonstrated an analogous result for the HSW capacity~\cite[pp. 555]{NC:2000}.

We have also proven that general POVM measurements are not needed to attain the  zero-error capacity of quantum channels, and that projective measurements are sufficient. These results were illustrated with an example of a quantum channel having a pentagon as characteristic graph.

Further work will include the study of relations with others areas of quantum information theory and quantum computation. More specifically, we think the theory of quantum zero-error is closely connected with quantum noiseless subsystems and the theory of graph states.

\section*{Acknowledgements}
The authors would like to thank the Programme AlBan, the European Union Programme of High Level Scholarships for Latin America (scholarship no. E05D051893BR), and the Brazilian National Council for Scientific and Technological Development (CNPq) (CT-INFO Quanta, grants \# 552254/02-9), for the financial support. This work has been partially supported by EC under project SECOQC (contract n. IST-2003-506813).

\section*{References}


\begin{thebibliography}{10}

\bibitem{Cover:91}
T.~M. Cover and J.~A. Thomas.
\newblock {\em Elements of Information Theory}.
\newblock John Wiley \& Sons Inc., New York, 1991.

\bibitem{BS:98}
C.~H. Bennett and P.~W. Shor.
\newblock Quantum information theory.
\newblock {\em IEEE Trans. Info. Theory}, 44(6):2724--2755, October 1998.

\bibitem{Sha:56}
C.~E. Shannon.
\newblock The zero error capaciy of a noisy channel.
\newblock {\em IRE Trans. Inform. Theory}, IT-2(3):8--19, 1956.

\bibitem{KoOr:98}
J.~K\"oner and A.~Orlitsky.
\newblock Zero-error information theory.
\newblock {\em IEEE Trans. Info. Theory}, 44(6):2207--2229, 1998.

\bibitem{MeAs:05a}
R.~A.~C. Medeiros and F.~M. de~Assis.
\newblock Quantum zero-error capacity.
\newblock {\em Int. J. Quant. Inf.}, 3(1):135--139, 2005.

\bibitem{MeAs:04d}
R.~A.~C. Medeiros and F.~M. de~Assis.
\newblock Quantum zero-error capacity and {HSW} capacity.
\newblock In {\em Proceedings of the The Seventh International Conference on
  Quantum Communication, Measurement and Computing QCMC'04}, volume 734 of {\em
  AIP Conference Proceedings}, pages 52--54. American Institute of Physics,
  2004.

\bibitem{ZR97c}
P.~Zanardi and M.~Rasetti.
\newblock Noiseless quantum codes.
\newblock {\em Phys. Rev. Lett.}, 79:3306, 1997.

\bibitem{Zan01}
P.~Zanardi.
\newblock Stabilizing quantum information.
\newblock {\em Phys. Rev. A}, 63:12301, 2001.

\bibitem{CK:05}
M.D. Choi and D.W. Kribs.
\newblock A method to find quantum noiseless subsystems.
\newblock {\em quant-ph/0507213}, 2005.

\bibitem{KBLW01a}
J.~Kempe, D.~Bacon, D.~A. Lidar, and K.~B. Whaley.
\newblock Theory of decoherence-free fault-tolerant universal quantum
  computation.
\newblock {\em Phys. Rev. A}, 63:042307, 2001.

\bibitem{HEB04}
M.~Hein, J.~Eisert, and H.~J. Briegel.
\newblock Multiparty entanglement in graph states.
\newblock {\em Phys. Rev. A}, 69:062311, 2004.

\bibitem{Bollo:98}
B.~Bollob\'{a}s.
\newblock {\em Modern graph theory}.
\newblock Springer-Verlag New York, Inc., New York, 1998.

\bibitem{Hol:98}
A.~S. Holevo.
\newblock The capacity of the quantum channel with general signal states.
\newblock {\em IEEE Trans. Info. Theory}, 44(1):269--273, 1998.

\bibitem{SW:97}
B.~Schumacher and M.~D. Westmoreland.
\newblock Sending classical information via noisy quantum channels.
\newblock {\em Phys. Rev. A}, 56(1):131--138, 1997.

\bibitem{NC:2000}
M.~A. Nielsen and I.~L. Chuang.
\newblock {\em Quantum Computation and Quantum Information}.
\newblock Cambridge University Press, Cambridge, 2000.

\bibitem{Lov:79}
L.~Lov\'{a}sz.
\newblock On the {S}hannon capacity of a graph.
\newblock {\em IEEE Trans. Info. Theory}, 25(1):1--7, 1979.

\end{thebibliography}

\end{document}